\documentclass[10pt, conference, letterpaper]{IEEEtran}

\usepackage[english]{babel}
\usepackage{url}
\usepackage{cite}
\usepackage{amsmath,amssymb,amsfonts}
\usepackage{algorithmic}
\usepackage{graphicx}
\usepackage{textcomp}
\usepackage{xcolor}
\usepackage{todonotes}
\usepackage{caption}
\usepackage{subcaption}

\def\BibTeX{{\rm B\kern-.05em{\sc i\kern-.025em b}\kern-.08em
    T\kern-.1667em\lower.7ex\hbox{E}\kern-.125emX}}

\usepackage{balance}

\begin{document}
\title{If Multicast is the Answer - What was the Question?}


\author{\IEEEauthorblockN{Dirk Trossen}
\IEEEauthorblockA{\textit{Huawei Research} \\
Munich, Germany \\
dirk.trossen@huawei.com}
\and
\IEEEauthorblockN{Jon Crowcroft}
\IEEEauthorblockA{\textit{University of Cambridge} \\
Cambridge, UK \\
Jon.Crowcroft@cl.cam.ac.uk}
}

\maketitle

\begin{abstract}
Multicast is (almost) as old as the Internet, having become a tool for increasing network efficiency but also enabling destination discovery in a number of key use cases, although misaligned economic interests have limited its deployment to domain-local usages. But recent advances in multicast technologies as well as the identification of new use cases for which IP multicast may be ill fitted yet network-level support may be desirable motivate to re-think old perceptions of multicast and its use in the Internet overall. For this, we return to the original question to which multicast is seemingly the right answer, based on which we outline emerging new answers to what multicast intends to achieve. Key to this is to re-formulate the multicast question in an attempt to semantically and architecturally align different answers, opening opportunities for more use cases to be served through multicast solutions, thus also driving the need for more research in this space. Our paper poses this new vision for multicast and investigates the alignment of existing and emerging multicast solutions with it, leading us to formulate a path for future research.
\end{abstract}

\begin{IEEEkeywords}
multicast, information-centric networking
\end{IEEEkeywords}

\section{From the Answer...}
\label{sec:answer}

Although the Internet initially started with unicast and broadcast communication semantics, it soon became apparent that selectively sending to more than one but not all Internet endpoints may be useful. What is often referred to as the \textit{Deering model} \cite{multicast} established a \textit{channel-based multicast} notion, denoted as $(S,G)$ with $S$ being the set of one or more senders and $G$ representing the identifier of the channel. \textit{Any source multicast}, i.e., $(*,G)$, allows anybody to send on the channel, akin to an Ethernet medium and useful for collaborative, interactive scenarios albeit less so in the wide-area Internet where lacking trust into the possible sender makes such semantic less useful and even dangerous. Realizing dissemination of information from curated, often authenticated sources, \textit{single source multicast} limits $S$ to a single sender and is often used for use cases such as IPTV, Internet radio, SW updates, and similar. Here, the notion of `being there' by gathering around a single source of information, akin to TV broadcast for, e.g., sport events, was a key driving use case, often directly driving social interactions after the event experience (often referred to as the `water cooler effect').

As outlined in \cite{Deering1990}, two key motivations exist for this channel semantic, namely (i) the increased \textit{efficiency} to deliver content over this channel to more than one receiver in the network and (ii) the possible \textit{discovery} of previously unknown destinations for future communication where the channel functions as a rendezvous mechanism.

In the Internet, Deering's channel model has been realized in what is known as \textit{IP multicast}, where dedicated IP addresses represent `Internet-wide' network interfaces \cite{RFC1112} to which receivers explicitly attach to receive any packets send to them. A group membership protocol \cite{RFC2236}\cite{RFC5186} enables not just localized multicast, e.g., in Ethernet networks, but building spanning trees where localized replication information is maintained in network-based \textit{branching points}, e.g., in protocol-independent multicast (PIM) \cite{RFC4608}.

However, most multicast usages in the Internet have been limited to single domains due to misaligned economic incentives between those benefitting from multicast (access network providers) and those providing inter-domain connectivity (peering providers), as explored in the seminal paper from Diot et al. \cite{multicast_incentives}. Crowcroft \cite{ICNMULTICAST} outlines other issues, such as Denial-of-Service aspects, a discussion we refer the interested readers for more insights. Nonetheless, multicast has played its role in realizing communication scenarios more efficiently.

One key observation in the realization of IP multicast is the intertwining of the communication semantic (the channel model) with that of selecting the endpoints to send (multicast) packets to (through maintaining localized group memberships for in-network replication of packets destined to a specific IP multicast address). This has limited use cases to those with relatively long-lived relations, expressed in those localized multicast (replication) groups.

For our revised notion of multicast, we thus return to formulating the right question for which multicast may be an answer. One starting point for this will be the decoupling of the communication from the endpoint selection semantic, as we will discuss next. We will then outline our vision to consolidate existing and emerging multicast solutions in Section \ref{sec:consolidation}, followed by examples in Section \ref{sec:example_answers}. We discuss why multicast matters again in Section \ref{sec:use_cases} albeit without falling into the same trap again as past multicast, as discussed in Section \ref{sec:economics}. We conclude our paper in Section \ref{sec:conclusion} with aspects that require invigorated research by the wider community.

\section{...Back to the Question}
\label{sec:question}

When returning to the question of what multicast may be, we argue that it is important to decouple the WHAT is being communicated from HOW it is delivered, particularly in relation to selecting the endpoints that ought to receive the multicast packets.

As for the WHAT, it is furthermore important to consider what is being identified when requesting a multicast semantic from the network, aligning with a similar question on 'What are the things that are identified by the identifiers?', posed in a recent IETF RTG WG interim meeting on evolving routing in the future Internet \cite{rtgwg_interim}.

Deering's model identifies a channel of communication, represented through a network-wide network interface and expressed through special IP addresses to which endpoints explicitly attach to by using the group membership protocol. Efforts like information-centric networking (ICN) \cite{CCN} instead identify content pieces \cite{RFC8609}. Proposals such as those developed in \cite{POINT} operate, similar to content delivery networks (CDNs), at the level of URLs, delivering responses possibly as multicast to more than one client.

But not just the answer to the question on WHAT is being requested in the communication has evolved, so has the answer to HOW the communication should be realized. Here, IP multicast, as suggested by Deering, foresees a strong coupling of time and space through its group model, requiring a common interest at the time of joining a group (e.g., as expressed by wanting to watch a common TV channel). This makes multicast communication \textit{intentional}, both through the definition of its (multicast) address and the explicit joining procedure for receiving packets, be it for the aforementioned delivery efficiency reasons or the purpose of discovery previously unknown destinations for future communication. It also positions its semantic separate from unicast communication where the attachment to the network (and thus the acquisition of an IP unicast) address will have a single destination receive anything sent to its unicast address.

By decoupling space and time, Van Jacobson presented in 2006 his initial proposition for ICN \cite{CCN}, including the use of ephemeral storage in the form of caches to allow delivering content at different times of the original dissemination of the content into the network. Through the decoupling, the occurrence of multicast delivery is \textit{coincidental} rather than intentional, thus not separating a possible unicast delivery from a multicast one; the former merely indicating the absence of the (coincidental) existence of another party interested in the requested content.

\textbf{Takeway:} When returning to a possible question for which (different realizations of) multicast may be the answer, separating the WHAT (semantic of communication) from the HOW (semantic for selecting the participating multicast endpoints) is fundamental.

We assert that this separation will allow us to formulate a joint semantic view on multicast that includes intentional models like IP multicast and those utilizing the coincidental existence of common interest in communication at the architectural level, while including its specific forms of delivery mechanisms under that same view.

\section{Vision to Consolidate Existing and Emerging Multicast}
\label{sec:consolidation}

Based on our main takeaway from the previous section, let us now outline a vision for multicast that is built upon the separation of WHAT and HOW, leading us to refactor the main components for a multicast system through explicit concepts that can then be jointly realized through specific solutions.

\begin{figure}[htbp]
  \centering
  \includegraphics[width=\linewidth,keepaspectratio]{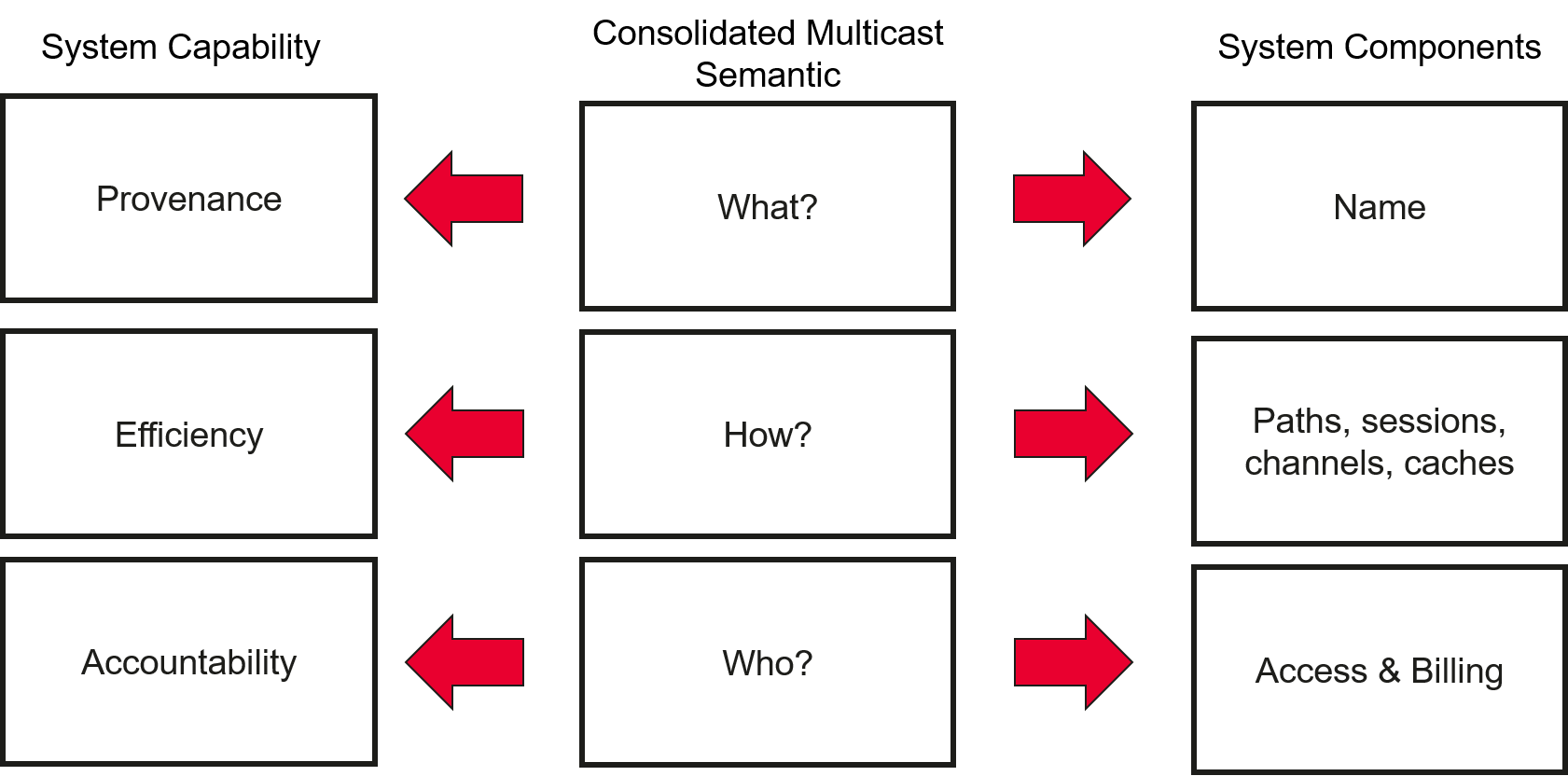}
  \caption{Vision for Refactoring Communication Systems}
  \label{fig:refactoring}
\end{figure}

As we can see in Figure \ref{fig:refactoring}, we extend the key WHAT and HOW questions with an addition WHO aspect. Through these, we achieve three key system capabilities, namely \textit{provenance}, \textit{efficiency}, and \textit{accountability}. Provenance is  realized through the \textbf{name} that links the multicast relation to its owner, possibly delegated to its distributor. Efficiency is achieved through realizations of the HOW question through methods like \textbf{paths}, \textbf{sessions}, \textbf{channels} but also \textbf{caches}, allowing to decouple time/space through moving from intentional NOW delivery to coincidental delivery over time (and space). Accountability provides the ability to not join authenticate \textbf{access} by those who may want to consume multicast content, but also creates a suitable flow of money that allows for \textbf{billing} for the transfer and storage of bits along the way from source to receiver(s).

Complementing our refactored communication system, we also define a consolidated multicast semantic as

\textit{A datagram with source address $S$ towards destinations $D_1$, ..., $D_n$, which in turn are identified through destination information $D$, is formed as one or more responses to adequate requests from $D_1$,..., $D_n$ towards S, where the ephemeral channel $C_M$ is defined through an identifying characteristic across all requests from $D_1$,..., $D_n$.}

In our semantic, the source address $S$ and destinations $D_1$, ..., $D_n$ represent the WHO in our vision, while the ephemeral channel $C_M$ allows for ensuring the provenance of the information that is being requested, while the delivery information $D$ is specific to the solution realizing the (efficient) delivery of the information.

We can see that our semantic preserves the channel model for intentional multicast communication, here in the form of the 'identifying characteristic' $C_M$, where said channel identifier represents a solution-specific parameter used to distinguish among different responses following the requests from any of the $D_1$,..., $D_n$ towards $S$. However, the receiver-driven aspect in our semantic also enables the coincidental notion of multicast where the ephemeral nature may be that of a single packet (in response to a declared interest in it), while aspects of caching is being represented by different sources $S$, possibly not even explicitly being identified through individual source addresses but merely identified by the ephemeral channel $C_M$.

Atop such communication semantic for multicast, reliability and application-specific sharing mechanisms may be implemented. Examples here are those in \cite{wb_multicast,ste_multicast}, which realize a content-oriented reliability mechanism through the multi-source multicast relation provided by the basic channel model realized through IP multicast.

\subsection{Components of the Resulting Communication System}
\label{sec:consolidation:impact}

While the next section studies the impact of the refactored communication system in Figure \ref{fig:refactoring} and our consolidated semantic on methods for realizing multicast empirically based on reviewing a set of solutions, let us discuss briefly what chief components of any realization need a closer look.

Firstly, the aspect of \textbf{naming} is crucial, i.e., what is being named and how is linkage to the originator or (authorized) distributor of information is being done. While IP multicast adapted much of its unicast sibling, we will see in the next section that new architectural approaches broke with this link, primarily to provide a new answer to the key naming aspect we outlined in Figure \ref{fig:refactoring}.

Secondly, the aspect of \textbf{delivery} is key, i.e., HOW is the actual information being disseminated across the network to reach all of those who signalled to receive it? We will see from our examples that the range of solutions span from routing-based approaches with in-network, channel-specific state, over path-based solution with centralized path-computation elements to construct the delivery relations to purely source-based methods that carry all necessary channel-specific state in the network with a minimal network state left to reduce costs.

Lastly, \textbf{accountability} has seen entry into multicast solutions from early days on, albeit largely provided through control plane solutions that construct suitable relationships being used for both authentication and billing. Again, newer architectural approaches that changed the focus of what is being named enabled accountability to be realized through linking names into the WHO is sending, while still leaving the billing aspect to a separate control plane.

In the following section, we shed more light on these aspects by studying various multicast solutions in more detail.

\section{Possible Different Answers}
\label{sec:example_answers}

In the following, we discuss how available multicast solutions map onto our semantics, providing thus specific answers to the same question of multicast, namely WHAT is being communicated and HOW it is delivered across the network to WHO ever requests it and can provide it.

\textbf{Original Internet multicast:} IP multicast \cite{multicast}, which is the most prevalent multicast realization in the Internet, realizes our proposed semantic by linking the channel identifier $C_M$ to an IP address within a dedicated assigned address space, i.e., an IP multicast address, which also serves as the destination address $D$. Following the abstraction of IP locators provided in IP unicast \cite{RFC2460}, $C_M=D$ represents a network-wide network attachment interface. Sending to this network-wide attachment point is achieved by first assigning $C_M$ to the sender $S$ \cite{RFC2375,RFC6308} for single source multicast, while any source multicast allows for any IP endpoint to act as source $S$. Each receiver explicitly joins the multicast reception through a group membership protocol \cite{RFC3376}, thereby realizing the explicit requests from $D_1$,..., $D_n$ towards $S$. The latter is used to build and maintain a spanning tree for delivery of multicast packets through explicit branching points along the delivery path, thus the requests may not be fully forwarded to $S$ but instead processed at a joint branching point in the network.

This procedure establishes network-internal state for branching packet delivery as the packet traverses the network from $S$ to $D_1$,..., $D_n$. Such state needs update for any change of the receiver set, be it through fluctuation of group membership itself or through mobility, i.e., the change of a network attachment point. Worthwhile noting is that the nature of the traffic is not explicitly captured in the group address and entirely left to the application to map onto the suitable IP multicast address by virtue of the aforementioned assignment procedure (and somehow conveying the assignment to the receivers for them join the multicast address reception). Furthermore, the single source multicast (SSM) model, outlined above, can also be extended to multiple source cases, each maintaining their own SSM relationship albeit joining the resulting trees in the network-internal branching points. With this, time and space may be decoupled, allowing for multi-source or selective replay scenarios to be realised.

We can see from this explicit process of address allocation and attachment procedure, in the form of the group membership protocol, that IP multicast communication indeed implements nn \textit{intention} to receive information sent to this (multicast) address, positioning IP multicast explicitly different from the unicast found in IP unicast packet delivery.

\textbf{Evolved routing architectures:} As discussed in Section \ref{sec:question}, decoupling time and space, including the usage of ephemeral caching in the network and in explicit caching infrastructures, shifts the intentional nature of IP multicast towards a coincidental one, where the coincidence lies in the joint interest in communication at the specific point in time as well as part of the network where joint delivery can be achieved. With this, the boundaries between multicast and unicast blur (intentionally - no pun intended).

This decoupling of time and space has been a key recognition both in research but also standards organizations. For instance, the IETF captured this aspect in recommendations for a routing architecture \cite{rfc6115} in 2015 as an outcome of the Routing Research Group in the IRTF, leading to developing solutions such as the Locator-Identifier Separation Protocol (LISP) \cite{RFC6830}\cite{LISP_INTRO}, where the endpoint identifier (EID) represents the WHAT with a possible multicast locator being used for HOW to deliver packets to the EID.

\textbf{Information-centric Networking:} Also research on ICN focussed on this key aspect of time/space decoupling, as discussed by Van Jacabson et al. in his seminal paper \cite{CCN}. Here, the ephemeral channel identifier $C_M$ may be that of a name for a piece of information that has previously been announced to the network by one or more sources $S$. HOW to select the endpoints to deliver the information, identified as $C_M$, differs across various realizations.

The most prominent ICN variant, called content-centric networking (CCN) as originally proposed in \cite{CCN}, utilizes a name-based routing model, where an \textit{interest} in content $C_M$ is explicitly expressed by one or more clients $D_1$,..., $D_n$, while the (name-based) routing infrastructure forwards those interest requests to one or more sources. Replies are sent as a \textit{data} packet, which is branched on the return path towards one or more of the receivers by using ephemeral state in the traversing network node, which has been built up during the forwarding of the initial interest packet(s). Hence, there is no explicit destination identifier $D$ since the return delivery merely relies on backtracking the original interest request. Also, unicast and multicast delivery purely differs in that sending a data packet to more than one requester becomes an aspect of synchronicity (and thus of the coincidence of the joint interest in content at this specific point in the network) of the incoming interest requests at the data-originating source, the direct result of the time/space decoupling objective.

Thus, the cost for multicast is represented by the additional packet copy operation along the return path from the sender to the (possibly more than) one receiver and the maintenance of the ephemeral state in the traversing nodes, plus the possible storage of information items in various sources $S$. With this, CCN decouples the WHAT is being identified from HOW endpoints are being selected by utilizing ephemeral forwarding information in traversing network nodes to remove the need for an explicit membership procedure, thus allowing for coincidental multicast to happen without it being intentionally declared as such.

The ICN realization in \cite{PURSUIT} pursues a different avenue, much alike the approach taken in the Locator/ID Separation Protocol (LISP) \cite{RFC6830}. Here, the requests from $D_1$,..., $D_n$ are not sent to the source but to an intermediary broker system (the rendezvous component in \cite{PURSUIT}, which is akin to the mapping service in LISP \cite{RFC6830}), which matches available sources with those requesting content, i.e., $D_1$,..., $D_n$. While LISP maps either onto IP unicast locators or an IP multicast address, PURSUIT uses a path computation element to construct suitable forwarding information, representing the destination identifier $D$ in our semantic, for the sources to send packets to $D_1$,..., $D_n$. The rendezvous/mapping protocol thus takes the place of the explicit group membership in IP multicast, while the WHAT in the form of the information name is separate from the information on which to send packets. The coincidental nature of multicast delivery here is defined by the ability of the separate broker system to form ephemeral relationships that may either be of unicast or multicast dleivery nature.

\textbf{Path-based variants of ICN:} Building on the path-based forwarding developed in \cite{PURSUIT}, derived from Bloom filter approaches \cite{LIPSIN,PSIRP}, the efforts in \cite{POINT} remove the need for an explicit membership procedure by focussing on use cases where application-specific requests of the same nature would lead to a possible multicast response. While the path computation and rendezvous aspects of \cite{PURSUIT} are preserved for unicast traffic, multicast responses are generated purely at the sender $S$ by combining the binary path information of individual incoming requests from $D_1$,..., $D_n$, resulting in an ephemeral forwarding information $D$ for the multicast return path. Basis for this capability is the representation of a network path through binary link information and its realization in commercially available SDN (software-defined networking) based switches, as described in \cite{PBF1}.

The efforts in \cite{POINT} come closest to source-created ephemeral information used for multicasting traffic that may result from incoming requests of the recipients. Key to the efficacy of the multicast transfer is the synchronicity of the incoming requests, directly impacting the possible size of the receiver set $D_1$,..., $D_n$. Furthermore, the ephemeral channel identifier $C_M$ is directly determined by the incoming requests and may be represented as, e.g., a hash of service-level information such as a service URL. With this, coincidental delivery (here from the originating source to the set of receivers) is defined through the reception of incoming requests for content within a given time interval; \cite{POINT} refers to this as the \textit{catchment interval}, which is determined based on the application semantics utilizing the communication to the destinations. In other words, application/service-level information plays a further role in impacting the information on the HOW the multicast response is forwarded to the recipients. For instance, chunk lengths in file retrieval scenarios were used to adjust the timing in the servers waiting for incoming client requests, thus increasing the likelihood for receiving more clients that happen to retrieve the same content.

\textbf{Path-based, source routed multicast:} There exist other binary path-based forwarding approaches beyond those in \cite{PBF1}. The Bit Indexed Explicit Replication (BIER) technology developed in the IETF \cite{RFC8279}, for instance, uses binary forwarding information to provide a L2/L2.5 underlay for multi-site connectivity. MSR6 (multicast source routing) \cite{MSR6}, a very recent IETF activity, proposes a recursive binary tree structure. Common to both approaches is the ability for the source to directly build the required forwarding information, thus forming the ephemeral relations envisioned by our multicast semantic, as captured in \cite{multicast_FRRM}.

\textbf{Looking across those answers:} When looking at the advances in multicast technologies, exemplified in the previous paragraphs, we can see that keeping network state as independent as possible from WHAT is being communicated is a hallmark of approaches that allow for building relationships directly at the sender with no semantic-specific state in the network at all, thus removing any need for adapting said state upon changes in receiver sets.

IP multicast suffers most from needing to maintain network state, while also CCN requires network state updates (in the form of name-based routing entries) when relationships between senders and receivers of information changes. Key obstacles in both approaches is the need for ephemeral network state that is dependent on the WHAT is being communicated, e.g., in the form of multicast address specific branching information or pending interest data to track the return of data packets over suitable network interfaces.

Source routing approaches, such as those in \cite{PURSUIT,POINT,RFC8279,multicast_FRRM}, minimize state in the network, limiting it purely to relationships between topological elements but not including any WHAT-related information. Binary representations are then used to form those ephemeral relationships, i.e., the HOW, with the set of receivers being entirely determined at the source itself at runtime. Through this, ephemeral relations can be much shorter in time, down to individual forward path requests, leading to possibly different multicast responses for any set of such forward requests.

\textit{As the takeaway from our examples, we observe that our consolidated semantic can be realized by many network-level multicast solutions albeit showing different capabilities in supporting ephemeral relationships, as expressed by the WHAT is being communicated, through the realization of HOW endpoint relations are being created; importantly, both intentional and coincidental forms of multicast delivery may occur.}

\section{Why Does Multicast Matter (Again)?}
\label{sec:use_cases}

So why does our discussion on a revised multicast semantic really matter? For an answer, we best look at possible use cases that benefit from it, particularly when considering multicast transmission in situations where commonly established solutions, such as IP multicast, do not provide suitable answers.

\textbf{Channel transmission:} Channels have been used as an abstraction for services for ages, with sources pushing data into them and sinks retrieving them. Receivers explicitly join into channels, such as explicitly tuning into the frequency (as for radio-based channel multicast) or implicitly doing so by changing channels on a TV set (which in turn will select the frequency assigned to the chosen channel). Broadcast services, such as television, radio, but also dissemination services, such as regular SW updates, use this dissemination model. Key here is the sender control, where delivery of content over the channel usually follows a schedule of delivery, e.g., a TV program, which aligns all clients to consuming the delivered content at the same time: space and time are tightly coupled for those use cases, directly impacting the user's experience of them. From a consumer perspective, the experience of gathering around a key event, such as a sporting event or the expression of national mourning (as in the UK at the time of writing this paper), drives the usage, complemented by the ability to also view those events at a later stage, utilizing the (possible) usage of multicast described next.

\textbf{Information retrieval:} Delivery of TV content through channel transmission has been accompanied through another delivery model, namely that of retrieving chunks of information, which combined constitute a timeseries of content, e.g., the TV experience. This chunking of a larger piece into smaller ones is further used for SW downloads but also (distributed) file storage. The key observation here is that the retrieval may not be continuous and its delivery does not necessarily follow a schedule set by the source but is instead periodically initiated by the receiver. Thus, it is particularly useful for retrieving stored catalogue content, as it happens in today's over-the-top video delivery platforms. The tight coupling of time/space in IP multicast has led to realizing delivery in these use cases entirely through source-based unicast replication in IP-based systems. Key problem here is the lack of a stable multicast group over a longer period; instead, relationships may change from one chunk request to another and even requests to the same chunk may be so time-varied that a single multicast group will not accommodate its delivery; instead, possible multicast relations are ephemeral, possibly down to single chunk relations. Such ephemeral relations, however, are supported when applying technologies such as ICN (in all its flavours) or its path-based derivates \cite{POINT} due to their time/space decoupling of delivery relations. This also allows for replicating content and delivering it through sources other than the original one, as explained before. HTTP-based realizations, e.g., for over-the-top video services, accomplish this through utilizing content delivery networks (CDN), acting as system-wide caches to improve efficiency, to move the unicast replication closer to the end user.

\textbf{Interactive group experiences:} From the beginning of IP multicast development, realizing small-scale group experiences was a key use case, facilitated by tools such as \cite{ste_multicast,wb_multicast}, while also early ICN work \cite{VOCCN} showcased using the ICN abstraction for interactive voice transmission. Layered atop the channel abstraction of IP multicast, these platforms provided object-oriented, reliable transmission of relevant content, including voice, video and whiteboard/editing information.

Here, efficient delivery of content is key due to the significant bandwidth demands for audio and video, in particular. But also the discovery aspect of multicast, as suggested by Deering in \cite{Deering1990}, is important for realizing `presence', even topically grouped, in virtual interaction to encourage participation and indicate liveness (e.g., of a debate or discussion), thus allowing for new communication groups to emerge, akin to the `water cooler moments' in physical social interactions.

Particularly, recent homeworking and homeschooling due to the Corona pandemic has increased the importance of this use case. However, these tools have been largely displaced by platforms that rely on server infrastructures, such as those found for interactive video calling (e.g., Skype, Teams, Zoom, and many more) or groupware for education. Although the drive for a server-centric solution may be explained with the often inter-domain nature of human-to-human connectivity, many use cases, however, may well originate and terminate within the same domain and would thus perfectly work with any of the solutions outlined in Section \ref{sec:example_answers} without running into the inter-domain incentive problems observed in \cite{multicast_incentives}. This is particularly true when considering enterprise deployments where software-defined wide area network (SD-WAN) technologies establish a single enterprise network domain overlay across possibly many network operators. In those deployments, server-less, pure multicast-based solutions may greatly improve efficiency as well as latency, thus positively impacting end user experience.

\textbf{NetReduce:} Distributed learning, as an emerging group of networked scenarios, often exhibits multipoint communication patterns, called \textit{collective communication}. For instance, the typical \textit{Reduce} function in MPI (Message Passing Interface)\footnote{\url{https://en.wikipedia.org/wiki/Message_Passing_Interface}} sends a number of input values for reduction, e.g., through aggregation, to a central point, with its result taken as input for the next iteration - thus often needing to be distributed back to the input endpoints. Relationships here are inherently ephemeral, where chosen aggregation functions may highly depend on the nature of the input data, thus preventing the use of stable group relations and hence the use of group-based (e.g., IP) multicast. Instead, building those ephemeral relations at the source (here the aggregation function) after receiving the necessary input would lead to network-level multicast and thus reduced communication overhead.

\textbf{Large scale IoT:} The proliferation of the Internet-of-Things has also evolved relations towards dynamic groups. For instance, actions may not just be limited to static device types, e.g., switching lights in a smart city, but may also include triggering collective sensing, which in turn is aggregated towards some higher semantics. This is more aligned with the distributed learning capabilities discussed in the previous paragraph.  Multicast may be desirable due to the scale (in number of devices) of communication, while supporting the dynamic nature of selecting the right input devices for the specific semantic aggregation.

While those examples above are not exhaustive, they represent a trend that has widened the potential for network-level multicast from group-based channel models to \textbf{highly dynamic relationships} in collective communication scenarios. The latter have yet to experience the use of efficient network-level multicast to meet ever stricter bandwidth and latency requirements. With the emergence of AI-native communication capabilities in, e.g., 6G Networks, it is high time for the networking community to provide the needed network-level multicast solutions that will enable those new use cases at the right efficiency and, thus, cost point.

But the question on why multicast may matter now cannot be about new usages only but must include the significant waste that unicast replications (compared to in-network branching of packets) signify and thus the energy consumption they represent. As shown in \cite{BBC}, for instance, combining multicast-based delivery with end user based caching can significantly reduce content transmission costs (up to 97\% of transmission and 74\% of energy costs, according to \cite{BBC}). Such combination of multicast with localized caching can also be employed for a redesign of CDNs, as proposed in \cite{FCDN}, with similar significant gains in efficiency. Both of these examples show the opportunity in terms of increased efficiency and thus decreased energy footprint per transferred bit even for scenarios in which multicast had not considered before.

\section{Falling Into The Same Trap?}
\label{sec:economics}

While the previous sections outlined that there are both new solutions and emerging demand for those solutions, let us now discuss how to avoid the traps that have already been at the forefront of failing to deploy IP multicast at large. The statement paper in \cite{ICNMULTICAST} identifies three key obstacles for a successful deployment.

The first is that of multicast being seen as enabling a \textit{Distributed Denial of Service} (DDoS) attack through the multiplication effect that stems from the any source model in IP multicast. Having the ability for any source to send data over a multicast channel amplifies any malicious intent, all while the network supports such attack through increased efficiency compared to a unicast-based one.

The second obstacle relates to \textit{billing} for the data being transmitted, caused by the decoupling between senders and receivers. This cannot just be solved by group membership (to account for who is listening) but also needs to consider the topological aspects of aggregating data transfers through branching points in the network.

Lastly, the \textit{inter-domain economic disincentive}, mentioned in the introduction and presented in \cite{multicast_incentives}, points at the problem of decoupling topological information from recipients in the inter-domain case, thus failing to provide the economic incentive for aggregation in the form of branching points to domains further downstream; the economic chain of peering and its associated billing breaks down, preventing the use of multicast beyond a single domain.

\textit{If those obstacles persisted, isn't any new demand for multicast doomed to fail being materialized through new solutions?}

We argue this not being the case, though. One main aspect of this argument lies in the design approach also mentioned in \cite{ICNMULTICAST}, namely the receiver-driven design of any solution. Already part of the original IP multicast design \cite{RDLM}, it is also inherently part of ICN designs in their publish-subscribe model. Our proposed semantic embodies the receiver-driven design in its
explicit requests sent to the sender $S$ for the formation of the ephemeral multicast group.

This addresses not just the DDoS aspect but also the problem of billing in some of the proposed solutions in Section \ref{sec:example_answers}. Path-based approaches, for instance, can provide information on branching points along the network traversal to the receivers (which are known in any case due to the incoming requests being sent for the formation of the ephemeral group). Furthermore, multicast billing approaches such as those in \cite{Henderson2000ProtocolIM} may be useful to be revisited for their adaptation on the multicast delivery solutions presented here.

When it comes to the economic disincentive at the inter-domain level \cite{multicast_incentives}, we argue that for most of our new demands in Section \ref{sec:use_cases}, inter-domain operation at the level of the multicast forwarding fabric is not required. This may be either due to the demand occurring in a single operator domain, such as for dense 5G multicast scenarios realizing localized AR/VR experiences or due to content being served entirely in localized Points of Presence (PoPs) as is the case for much of the consumed content in the Internet today \cite{CISCO}, or due to the solution realizing a shim overlay that may span several (peering) network operators but still form a single \textit{multicast} domain with full knowledge and control over recipients and branching point alike; here, the BIER and MSR6 solutions discussed in Section \ref{sec:example_answers} are examples for such shim overlays.

\textit{The takeaway here is that the woes of previous multicast deployments can be overcome by observing the lessons learned from those deployments, as is done in our proposed new multicast semantic.}

\section{What Next?}
\label{sec:conclusion}

We have observed in our discussion that the notion of multicast is an evolving one, not being cemented by its only current deployment in the Internet. This shows the role for research in exploring new ways to efficiently implement collective communication patterns down to the network level in order to achieve the possible efficiency that inherently lies in the nature of multicast relations. Specifically with examples like ICN, there is a clear need but also benefit to rethink and question assumptions of existing systems, proposing architectures and solutions that address the ephemeral nature we can observe in many of those multicast relations.

We believe that this ephemeral nature is best captured by deeply re-thinking the purpose of communication (the WHAT) in relation to its realization (the HOW). This re-thinking will need to touch upon a number of aspects. For instance, \textit{addressing} needs to be rethought, as we can observe in new architectures like ICN where naming of content is partially, or in some flavours even entirely decoupled from the method to deliver it. The role of \textit{routing} needs careful consideration, as the management of network-wide state is a major problem that source-based approaches may reduce or even entirely remove, not only reducing costs for maintaining such state but also allowing for the dynamicity of multicast relations that newer use cases demand. \textit{Scalability} is another aspect, which was important in the original IP multicast design, but lost importance due to the lack of inter-domain deployment, as discussed in the introduction. However, even local, e.g., 5G, deployments may require significant scale in terms of network nodes and endpoints alike albeit in a more local setting.

Furthermore, the right \textit{architectural framework} needs formulation. From our insights in this paper, we can see the need for highly efficient yet scalable (in terms of supported endpoints and tree depths alike) domain-local multicast architectures, while the use of a  shim overlay atop those new multicast and existing unicast architectures may be the right approach to avoid the economic incentive trap of old times, through providing suitable billing interfaces from to the traversing transport networks to ensure that economic value flows aligned with the packets in the network. Key here is a thorough \textit{comparison} of the various technologies, not just in terms of technologies, their performance, e.g., in terms of scalability and costs, but also at an architectural level to avoid falling into the trap of failed deployment again, as discussed in the previous section.

Standardization organisations, such as the IETF (and its sister organisation, the research-oriented IRTF), are not standing still either. The aforementioned BIER and MSR6 works are placed in the IETF with solution specifications being developed right now. SDN as the basis for \cite{PBF1} has a direct standards and thus commercial basis, allowing for deploying solutions in existing networks. Efforts such as those on \textit{semantic routing} \cite{SEMANTICROUTINGINTRO} have recently raised the need for an architectural framework that would programmatically enable a range of communication semantics, including multicast \cite{SEMANTICROUTINGVISION} albeit with no solutions so far provided that can be deployed.

We conclude that the key to re-thinking of what multicast may be is to uncover the original question to it, namely WHAT do we want to communicate and HOW do we deliver it? We outlined a consolidated semantic around these separate but interrelated questions that covers a wide range of multicast solutions from stable, long-lived relations to extremely ephemeral, and short-lived ones experienced in newer use cases. Key to all those solutions as well as to our semantic is to preserve the key idea of multicast, namely to enable high efficiency in data delivery as well as discovery of future communication to come, all within an intentional or coincidental view of how multicast ultimately happens in the network. We hope that research will continue filling this semantic with suitable answers, leading to more network-level multicast in the future.

\section{Acknowledgements}
Useful comments and hints have come from Tony Li, Dino Farinacci, Lixia Zhang, and Luigi Iannone, who share no responsibility for the resulting text.

\bibliographystyle{ieeetr}
\bibliography{sigproc, rfc}

\end{document}